\input{aipcheck.tex}

\documentclass[article]{aipproc}

\layoutstyle{6x9}

\SetInternalRegister\hbadness{8000} 

\newcommand{\bea}{\begin{eqnarray}}
\newcommand{\eea}{\end{eqnarray}}       
  
\begin{document}

\title 
      [An Effective Theory for the Four-Body System]
{An Effective Theory for the Four-Body System}

\classification{43.35.Ei, 78.60.Mq}
\keywords{Document processing, Class file writing, \LaTeXe{}}

\author{L. Platter}{
  address={Forschungszentrum J{\"u}lich, Institut f{\"ur} Kernphysik,
    D-52425 J{\"u}lich, Germany},
  altaddress={Helmholtz-Institut f\"ur Strahlen- und Kernphysik (Theorie),
Universit\"at Bonn, Nu\ss allee 14-16, D-53115 Bonn, Germany}
  }

\copyrightyear  {2004}

\begin{abstract}
We consider the non-relativistic four-body system with large
scattering length and short-range interactions within an effective
theory with contact interactions only. We compute the binding
energies of the $^4$He tetramer and of $\alpha$-particle. The well-known
linear correlation between the three-body binding energies and the
four-body binding energies of these physical systems can be understood
as a consequence of the absence of a four-body force at leading order.
\end{abstract}

\date{\today}

\maketitle

{\it Introduction -}
Effective theories are particularly well suited to describe low-energy properties
of physical systems in a model-independent way. Results and errors can be
improved systematically and thus, effective theories can be used in principle
to compute observables to arbitrary high precision.
If the scattering length $a$ of two particles is much larger than the typical low-energy
length scale $\ell$ of the system, one can use an effective theory with contact
interactions only, to compute observables in an expansion in $\ell/a$.
The beautiful feature of this theory is, that it doesn't make any assumptions
about the underlying physics, besides that the resulting potential is short-ranged
and produces a large scattering length. This allows for a systematic comparison
of low-energy systems at different length scales. Particularly interesting, in
this respect, are few-body systems with large scattering length: For the three-body
system it turns out that a further piece of three-body information is needed
to fully describe observables. In the effective theory this is reflected by
a leading order three-body interaction which has to be renormalized accordingly.
Its renormalization group flow is governed by a limit cycle.  
In \cite{Platter:2004qn} we considered the four-boson system within this
framework. Here we will summarize the most important results without going
into the technical details and will also present results from the
four-nucleon sector as recently given in \cite{Platter:2004zs}.
\\

{\it Four-Boson System -}
At leading order the regulated effective low-energy 
potential generated by a non-relativistic effective field theory with
short-range interactions is given by
\begin{equation}
\langle {\bf u_1}|V|{\bf u'_1}\rangle=
\langle {\bf u_1}|g \rangle \lambda_2 \langle g|{\bf u'_1} \rangle ~,
\label{effpot_2}
\end{equation}
\begin{table}[t]
\begin{tabular}{|c||c|c||c|c|}
\hline
system & $B^{(0)}$ [mK] & $B^{(1)}$ [mK] & $B_{\rm BG}^{(0)}$ [mK] 
& $B_{\rm BG}^{(1)}$ [mK] \\ \hline\hline
$^4$He$_3$ & 127 & [2.186]   & 125.5  & 2.186\\
$^4$He$_4$ & 492 & 128  & 559.7  & 132.7\\ \hline
\end{tabular}
\caption{\label{tab:results}Binding energies of the $^4$He
trimer and tetramer in mK.
The two right columns show the results by Blume and Greene 
\cite{Blume:2000} (denoted by the index BG)
while the two left columns show our results. The number in brackets was
used as input to fix $\lambda_3$.}
\end{table}
where ${\bf u_1}$ and ${\bf u'_1}$ are the relative 
three-momenta of the incoming and outgoing particles,
 respectively. $g({\bf u_1})=\exp(-u_1^2/\Lambda^2)$ is a regulator function 
which suppresses momentum contributions with $u_1 \gg \Lambda$.
 The coupling constant $\lambda_2$ can be matched
to the scattering length or the two-body binding energy.
The three-body force (which is needed to renormalize the three-body
system) is given by
$
V_{3}=|\xi\rangle\lambda_3\langle\xi|~,
$
where $\xi({\bf u_1,u_2})=
\exp(-(u_1^2+{\textstyle \frac{3}{4}}u_2^2)/\Lambda^2)$ is the
corresponding regulator function and ${\bf u_2}$ represents the
second Jacobi momentum in the three-body system. The three-body
coupling constant $\lambda_3$ can be fixed by demanding that the
binding energy of the shallowest three-body bound state stays
constant as the cutoff $\Lambda$ is changed. This renormalization
 prescription leads to the limit cycle mentioned above.\\
\begin{figure}[b]
\centerline{\includegraphics*[width=2.5in,height=1.8in,angle=0]{tjon_40_2.eps}
\includegraphics*[width=2.5in,angle=0]{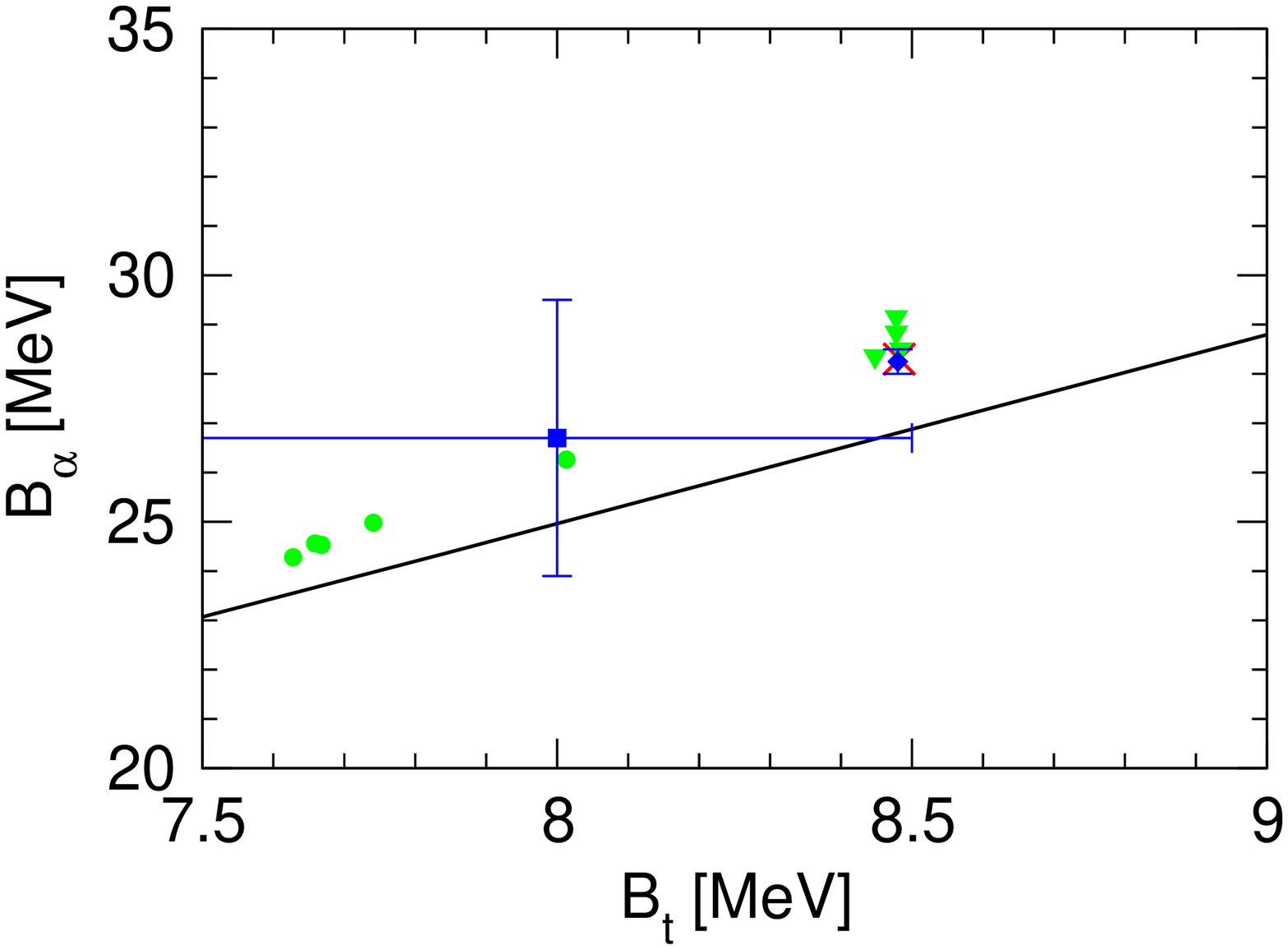}}
\caption{\label{fig:tjonline}
The left plot shows the correlation between
the ground state energies of the $^4$He trimer and tetramer. 
The solid line shows the leading order 
effective theory result and the cross denotes the calculation for
the LM2M2 potential by Blume and Greene \cite{Blume:2000}. The triangles
show the results for the TTY, HFD-B, and HFDHE2 potentials \cite{Lew97,Naka83}.
The right plot shows the correlation between the triton and the
$\alpha$-particle binding energies. The solid line shows our leading order
result using the singlet scattering length $a_S$ and the deuteron binding
energy $B_d$ as two-body input. The grey dots and triangles show various
calculations using phenomenlogical potentials without our including
three-nucleon forces, respectively \cite{Nogga:2000uu}.
 The squares show the results of
chiral EFT at NLO and N$^2$LO \cite{Epelbaum:2002vt,Epelbaum:2000mx} 
while the cross shows the experimental point. 
}
\end{figure}We have used the Yakubovsky equations to compute the binding energies
of the $^4$He$_4$ tetramer. As three-body input we used the energy
of the shallowest three-body bound state as calculated 
by Blume and Greene (BG) \cite{Blume:2000}. An analysis
of the cutoff dependence of the results for the binding energies
shows that no four-body force is needed to renormalize
the four-body sector.
Further, a comparison of our results with the values obtained by BG
\cite{Blume:2000} is shown in Table \ref{tab:results}.
The results of their calculation for the trimer and tetramer are given in the 
two right columns of Table~\ref{tab:results}, while our results are given in 
the two left columns.
In general, our results are in good agreement with the values of BG.
We also analyzed the correlation between three-body and four-body
binding energies.
The left plot in Fig.~\ref{fig:tjonline} shows the ground state
energy $B^{(0)}_4$  of the $^4$He tetramer as a function of the
 trimer ground state energy $B^{(0)}_3$. The solid line is the leading order 
result of our effective theory calculation and the cross denotes the result
of the calculation by BG for the LM2M2 potential \cite{Blume:2000}. 
For the ground states of the trimer and tetramer, calculations with
other $^4$He potentials are available and are shown as well~\cite{Lew97,Naka83}.
\\
\\
{\it Results for the Four-Nucleon System -}
It is straightforward to apply the above procedure to the four-nucleon system.
Without going into the technical details of the inclusion of spin and isospin
we present out results:
For the $\alpha$-particle we obtain a binding energy  $B_{\alpha}=26.9$ MeV
if we use the triplet scattering length, the deuteron binding energy and triton
binding energy as input.
By varying the three-body binding energy we are again able to observe a 
linear correlation between the three-body
and four-body binding energies, which is shown in the right plot of 
Fig. \ref{fig:tjonline}.  Thus, we conclude that the Tjon line
is a result of the large scattering lengths in the nucleon-nucleon system.
More information and further references on this topic can be found in
\cite{Platter:2004zs}.
\\
\\ 
{\it Summary -}
We have shown, that four-body systems with large scattering length can
be described within the framework of the effective theory with
contact interactions. Our results for the binding energies of the $^4$He
tetramer and the $\alpha$-particle are in good agreement with theoretical
calculations and the experimental value, respectively. Universal properties
 of these systems like the Tjon line turn out to be a result of the large
 scattering length in the two-body
sector. This can be understood in detail by the absence of a four-body force
 at leading order. As a consequence, all two-body interactions that produce
 a large scattering length in the two-body sector will give four-body
binding energies lying close to the Tjon line.\\
In the near future further effort should be devoted to the computation of
scattering observables and the inclusion of effective range corrections. 
\begin{theacknowledgments}
This work was done in collaboration with H.-W.~Hammer and U.-G.~Mei\ss ner. 
This research was supported in part by the the Deutsche Forschungsgemeinschaft
through funds provided to the SFB/TR~16.
\end{theacknowledgments}


\begin{thebibliography}{99999}
\bibitem{Platter:2004qn}
L.~Platter, H.~W.~Hammer and U.~G.~Meissner,
Phys.\ Rev.\ A (in press), arXiv:cond-mat/0404313.

\bibitem{Platter:2004zs}
L.~Platter, H.~W.~Hammer and U.~G.~Meissner,
arXiv:nucl-th/0409040.


\bibitem{Blume:2000}
D.~Blume and C.H.~Greene, J.\ Chem.\ Phys.\ {\bf 112}, 8053 (2000).

\bibitem{Lew97}M.\ Lewerenz, J.\ Chem.\ Phys.\ {\bf 106}, 4596 (1997).

\bibitem{Naka83}S.\ Nakaichi-Maeda and T.K.\ Lim, Phys.\ Rev.\ A {\bf 28}, 
  692 (1983).

\bibitem{Nogga:2000uu}
A.~Nogga, H.~Kamada and W.~Gl{\"o}ckle,
Phys.\ Rev.\ Lett.\  {\bf 85}, 944 (2000)
[arXiv:nucl-th/0004023].

\bibitem{Epelbaum:2002vt}
E.~Epelbaum, A.~Nogga, W.~Gl\"ockle, H.~Kamada, U.-G.~Mei\ss ner and H.~Witala,
Phys.\ Rev.\ C {\bf 66}, 064001 (2002)
[arXiv:nucl-th/0208023].


\bibitem{Epelbaum:2000mx}
E.~Epelbaum, H.~Kamada, A.~Nogga, H.~Witala, W.~Gl\"ockle and U.-G.~Mei\ss ner,
Phys.\ Rev.\ Lett.\  {\bf 86}, 4787 (2001)
[arXiv:nucl-th/0007057].

\end{thebibliography}
\end{document}